\documentclass{article}
\usepackage{spconf,amsmath,graphicx,hyperref}
\usepackage{xcolor}
\usepackage{booktabs}

\title{21Codec: Decoder-side Speaker Control for Text-to-Speech}

\name{David Lindevelt, Suzan Verberne, Joost Broekens}
\address{
     \textit{Leiden University}\\
     \textit{LIACS} \\
     Leiden, The Netherlands 
}

\begin{document}
\maketitle
\begin{abstract}
	We propose 21Codec, a streaming codec model with reduced recoverable speaker
	information from the token stream and decoder-side speaker control.
	Existing codec models infer speaker identity from token stream. Making them
	streamable is not sufficiently robust to speaker drift, because each chunk
	re-infers the identity from the local chunk context alone.
	Our adversarial speaker classifier and NANSY-style formant-shift data
	augmentation pushes the model to rely on the provided speaker embedding while
	reducing the recoverable speaker information from the token stream.
	We show that 21Codec can control speaker identity through the conditioning
	vector, and that the speaker identity is maintained across chunks during
	streaming.
	With the ability to control speaker identity even in mid-stream, 21Codec is a
	promising candidate for streaming TTS systems for better controllability of
	speaker identity.
\end{abstract}
\begin{keywords}
	text-to-speech, speaker information, audio compression, neural audio codec, finite scalar quantization
\end{keywords}
\section{Introduction} \label{sec:intro}
Autoregressive text-to-speech (TTS) systems increasingly operate over discrete
audio tokens, which makes the codec an intermediate representation for the
whole pipeline. Most of the existing codec models
\cite{julianFiniteScalarQuantization2025, jiWavTokenizerEfficientAcoustic2025,
	kumarHighFidelityAudioCompression2023a} are trained to reconstruct the audio
with a sufficiently large window, and their decoders exploit the global context
for inferring mostly static information such as speaker identity. Speaker
identity is inferred from the full token sequence rather than supplied
explicitly. When these codec models are used in streaming TTS
systems\cite{liuStyleStreamRealTimeZeroShot2026}, the decoder is forced to
operate on short chunks of tokens, which makes the speaker identity inference
unstable: When an unchanged token sequence is decoded in chunks, each chunk
re-infers the identity from the local chunk context alone, and arrives at a
different speaker identity. So, in current SOTA codec models, the decoded voice
audibly drifts between speaker identities, even within the same utterance. This
makes the models incompatible with the real-time synthesis where shorter token
windows are needed for better real-time performance.

We argue that this is a representational problem rather than a decoding
problem. The speaker information is encoded in the codec codes. We propose
21Codec\footnote{A complete text-to-speech system has two halves: an acoustic
	model and a codec vocoder. This paper is about the second half, therefore half
	of Douglas Adams' Answer to Life, the Universe, and Everything ($42$) }, an 800
bps streaming codec at 50 Hz frame rate that suppresses speaker identity from
the token stream. 21Codec uses an adversarial speaker classifier and
NANSY-style formant-shift data augmentation. It then reintroduces speaker
identify at the decoder through adaptive modulation on speaker embedding.
Identity enters once, as a single external embedding that conditions every
frame, instead of being re-inferred from each chunk, preventing speaker drift.

Conditioning allows our model to re-voice the same token sequence onto
different speakers without changing its content. Suppressing the speaker
identity in a learned representation and re-supplying it at the decoder is
established in voice
conversion~\cite{zhangNonParallelSequencetoSequenceVoice2020,
	chouMultitargetVoiceConversion2018}, but those systems are not streaming and
operate on an internal continuous latent. We operate on the discrete $0.80$
kbps bitstream the TTS pipeline uses, and decode it chunk-causally. No
component, parameter, or added latency enters the pipeline.

Our first motivation is to keep identity consistent throughout streaming
inference. Our second motivation concerns our prior observation about
\textit{where} speaker identity should enter a controllable TTS system. Style
control is far more sensitive to conditioning depth than speaker identity is,
yet compete for the same early layers. Under a joint objective, style should
hold those layers and identity should be applied as late as possible. We remove
identity from the token stream. So, the two signals never compete for the
capacity at any depth.

Our contributions are: \textbf{1.} A streaming codec model with reduced speaker
identity in its codes. Re-voicing a source utterance onto a different speaker
puts the output closer to the reference. \textbf{2.} The ability to change
speaker identity mid-stream, at no additional bitrate. \textbf{3.} An
evaluation showing that causal streaming decoding matches non-causal decoding
down to a chunk size of 500 ms (25 frames).

\section{Method} \label{sec:method}
\begin{figure}[t]
	\begin{minipage}[b]{1.0\linewidth}
		\centering
		\centerline{\includegraphics[width=\linewidth]{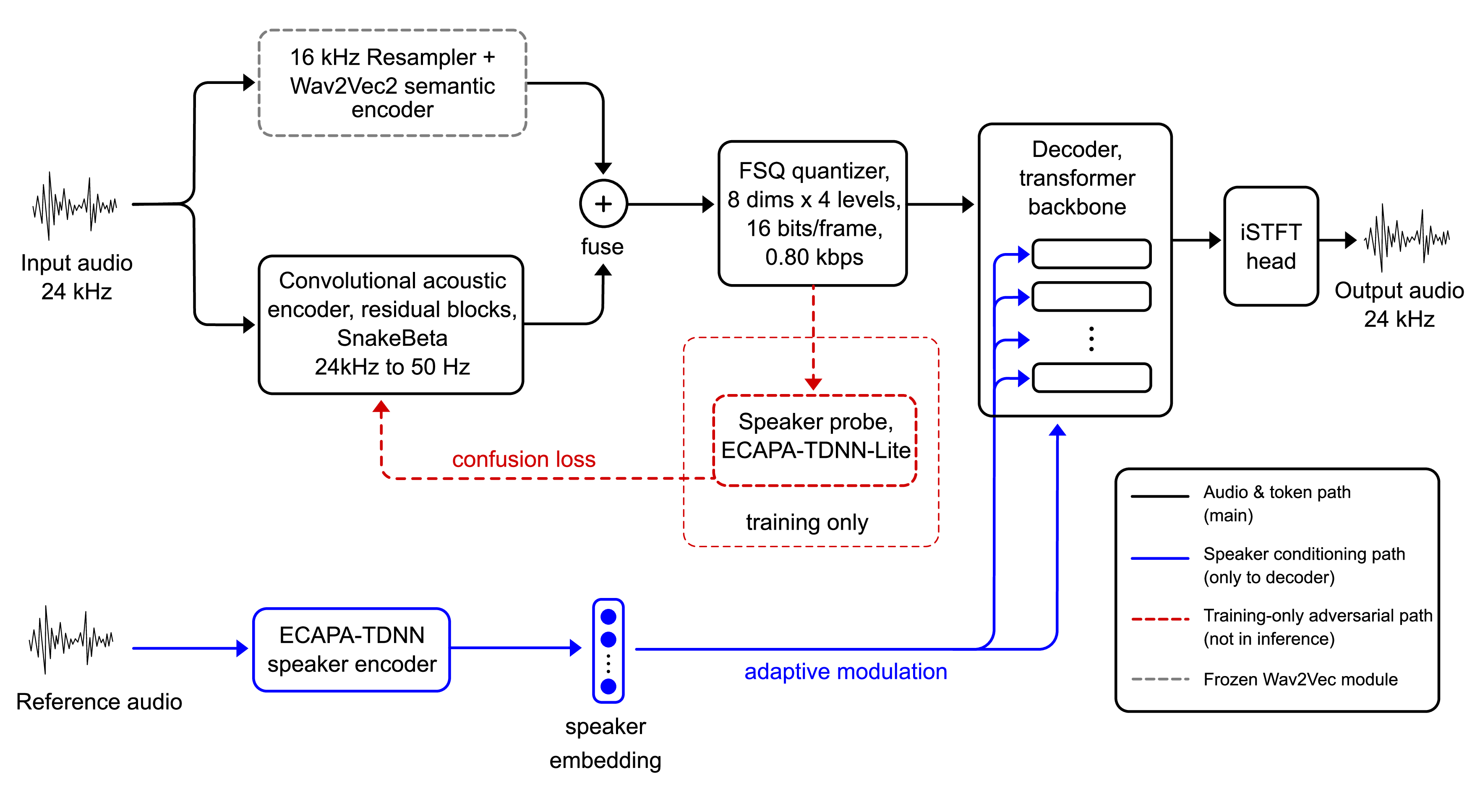}}
	\end{minipage}
	\caption{21Codec Architecture. The encoder is non-causal, the decoder chunk-causal.}
	\label{fig:architecture}
\end{figure}

\textbf{Our encoder} consists of acoustic and semantic encoders. The acoustic
encoder is a convolutional neural network with residual blocks with SnakeBeta
activation function. The acoustic encoder downsamples 24 kHz input audio to 50
Hz frame rate. The frozen
Wav2Vec2~\cite{communicationSeamlessMultilingualExpressive2023} semantic
encoder output is concatenated and fused with acoustic encoder output. The
fused representation is quantized with
FSQ~\cite{mentzerFiniteScalarQuantization2024}. With eight dimensions at four
levels each, every frame carries $\log_2 4^8 = 16$ bits, giving a bitrate of
$0.80$ kbps. We adopt FSQ~\cite{mentzerFiniteScalarQuantization2024} rather
than residual vector quantization to avoid codebook collapse. We measure
codebook utilization over 582 hours of LibriTTS dataset. All 65{,}536 implicit
codes are used, the code distribution reaches $15.66$ of $16$ bits of entropy
($783$ bps and $97.9\%$ of the theoretical maximum), and the total correlation
between the code dimension is $0.18$ bits. So, an ideal entropy coder could
recover only $0.02$ kbps, which shows the codes are well utilized and close to
incompressable.

\textbf{The decoder} consists of a transformer backbone and ISTFT
head~\cite{siuzdakVocosClosingGap2024, kanekoISTFTNETFastLightweight2022a} to
reconstruct the audio from quantized representation. While our encoder is
non-causal, our decoder is chunk causal. Self-attention in transformers creates
a mask to block only future chunks. It is designed to work with random chunk
sizes which makes the model streamable at arbitrary chunk sizes for TTS
systems. We use DiT~\cite{peeblesScalableDiffusionModels2023} inspired adaptive
modulation in decoder to condition on
Ecapa-TDNN~\cite{desplanquesECAPATDNNEmphasizedChannel2020} speaker embeddings,
as in \cite{mustafaStyleMelGANEfficientHighFidelity2021,
	paulSpeakerConditionalWaveRNN2020a}, but on a representation from which
identity has already been suppressed. Separately, we implement
ECAPA-TDNN-Lite~\cite{liLightweightApplicationsAsymmetric2022} speaker
classifier probe $\phi$. The probe $\phi$ is trained to predict the speaker
from the frozen codec's quantized codes $q$ under an additive-angular-margin
softmax objective. It is trained on the same formant-shifted distribution,
always under the original speaker label, so it becomes shift-invariant and
relies on cues the augmentation cannot remove. The codec is then trained
against it with a confusion penalty that drives the probe's posterior toward
uniform, $\mathcal{L}_{\mathrm{conf}} = D_{\mathrm{KL}}\!\left(\mathcal{U}
	\;\|\; p_\phi(s \mid q)\right)$, evaluated on margin-free cosine logits, with
gradients flowing back into the encoder. The two are optimized in alternation.
The penalty is label-free and therefore applies to every utterance. The full
generator objective is $\mathcal{L}_G = 45\mathcal{L}_{\mathrm{mel}} +
	\mathcal{L}_{\mathrm{energy}} + 3\mathcal{L}_{\mathrm{adv}} +
	6\mathcal{L}_{\mathrm{feat}} +
	\lambda_{\mathrm{spk}}\mathcal{L}_{\mathrm{conf}}$ with $\lambda_{\mathrm{spk}}
	= 1$.

We use \textbf{NANSY~\cite{choiNeuralAnalysisSynthesis2021a} style data
	augmentation} to the input audio. The module warps the spectral envelope along
the frequency axis by a random factor between $1.14$ and $1.33$ while leaving
the harmonic fine structure intact. This way, the fundamental frequency (F0)
and the intonation contours are preserved, while the speaker timbre (formants)
changes. We apply the formant shift augmentation and measured the speaker
embedding shift using cosine similarity between the original and augmented
audio. The random factor boundaries are chosen based on our analysis of that
shift.

\section{Experimental Setup} \label{sec:setup}

\textbf{Models}. We compare 4 models: 21Codec, 21Codec-U,
NeuCodec~\cite{julianFiniteScalarQuantization2025}, and
WavTokenizer~\cite{jiWavTokenizerEfficientAcoustic2025}. 21Codec is both
streamable and speaker-conditioned. For better comparison, we implement
21Codec-U: an unconditional version with the same architecture; without the
speaker adversarial and formant-shift augmentation.

\textbf{Training details.} Our models are trained on single node with 4 NVIDIA
H100 GPUs for $560k$ iterations. Training uses $6$ seconds of audio slices with
effective batch size $128$. Shorter samples are looped to $6$ seconds. Models
are optimized using AdamW. For codec, discriminator, and speaker adversarial
learning rates $4\times 10^{-4}$, $2\times 10^{-4}$, $1\times 10^{-3}$ and
cosine annealing schedulers with $10k$, $10k$, $2k$ warmup steps respectively.
We adapt zclip~\cite{kumarZClipAdaptiveSpike2025} for per-module adaptive
gradient clipping using EMA and z-score outlier detection. We will release our
code repository.\footnote{\url{https://github.com/dcze4/21Codec}}

Training data is a weighted mixture of seven corpora
(Table~\ref{tab:train_data}). Groups are sampled at fixed weights rather than
in proportion to their sizes, so the distribution the model sees is set by the
weights, not by corpus hours. Utterances shorter than $4$\,s or longer than
$30$\,s are filtered out. Audio is converted to $24$\,kHz mono and
loudness-normalized to EBU R128 at $-22$\,LUFS integrated, $7$\,LU loudness
range and a true peak of $-3$\,dBTP. Common Voice and HUI are additionally
passed through a source separation, de-reverberation and denoising chain before
normalization; remaining corpora did not.

\begin{table}[t]
	\centering \small
	\caption{Training mixture. Weights are per-batch sampling probabilities, not proportions of the available hours.}
	\label{tab:train_data}
	\begin{tabular}{llrrc}
		\toprule
		Corpus                    & Lang         & Hours   & Spk.  & Weight \\
		\midrule
		In-house (audiobooks)     & en           & 862.08  & 44    & .30    \\
		LibriTTS-R                & en           & 449.68  & 2277  & .18    \\
		Common Voice 23.0         & 54           & 1865.16 & 36382 & .15    \\
		MLS                       & \phantom{0}7 & 5870.32 & 518   & .15    \\
		In-house (acted dialogue) & \phantom{0}2 & 532.37  & 2135  & .15    \\
		HUI                       & de           & 322.88  & 121   & .05    \\
		Expresso                  & en           & 4.87    & 4     & .02    \\
		\bottomrule
		Total                     & 54           & 9907.36 & 41481 & 1.00   \\
	\end{tabular}
\end{table}

\textbf{Evaluation details.} We evaluate identity control (re-voicing, SMOS),
recoverability (speaker probe), stability (speaker drift, chunk size), and cost
(reconstruction, MOS). \textit{Identity control.} Training data is
English-dominant, so we evaluate on every unseen LibriTTS-R speaker with at
least two utterances of $15$\,s or longer, total of $80$ in speakers
(\textit{EN-80}), sampling 2 utterances per speaker. Re-voicing trials pair
each source with a reference utterance from the same speaker and four from
different speakers. By alternating the source and reference in the same-speaker
condition, we have $480$ trials. Similarity listening test instead uses five
languages (en, nl, de, fr, it) since judging whether two voices match does not
require understanding the content. It consists disjoint set of $100$ source and
$100$ reference speakers, balanced over languages, and presents the re-voiced
output next to a codec reconstruction of either the source or the reference
speaker. Each source is paired with one matched-language and four
cross-language references, plus two calibration conditions: a genuine same- and
different-speaker pairs.

\textit{Recoverability.} Speaker probe evaluation uses $5000$ utterances from
speakers seen during training, $355$ of them are held-out. The probe is trained
on the frozen codes of both unconditional baseline and 21Codec for $100k$ steps
using the same training setup used in codec training. Then, we report the top-1
accuracy. \textit{Stability.} We select $40$ utterances across five languages,
each 20--30 seconds and from a different unseen speaker. These utterances are
used to measure speaker identity drift when the same token sequence is decoded
in various chunk sizes. Speaker embeddings are extracted from non-overlapping 2
second windows, and the cosine similarity is measured between synthesized and
source utterance windows. The chunk-size sweep evaluation uses $30$ utterances
from $30$ different LibriTTS-R speakers. \textit{Cost.} Reconstruction
evaluation uses EN-80 set, total 160 trials. MOS evaluation uses the same
subset as chunk-size evaluation.

\textbf{Evaluation metrics.} We evaluate the models on objective metrics:
DNSMOS ovrl\cite{reddyDnsmosP835Nonintrusive2022}, NISQA
mos\cite{mittagNISQADeepCNNselfattention2021}, WER/CER and speaker similarity
using cosine similarity. Reference-based metrics such as PESQ, STOI and SI-SNR
are not reported because they measure waveform mismatch rather than perceived
quality for every system compared here.
Whisper~\cite{radfordRobustSpeechRecognition2023} turbo is used as ASR model
for WER/CER. For speaker similarity, we use
ECAPA-TDNN~\cite{desplanquesECAPATDNNEmphasizedChannel2020} speaker embedding
model to extract embeddings from the source, reference and synthesized audio,
and compute cosine similarity between them.

\textbf{Listening Tests.} We conducted two human evaluation studies: \textit{1.
	SMOS}: we asked $21$ participants to rate $30$ samples on a $1$ to $5$ scale,
where $1$ means different speaker and $5$ means same speaker. \textit{2. MOS}:
we asked $24$ raters to rate the naturalness of the audio on a $1$ to $5$
scale, where $1$ means bad and $5$ means excellent. For both studies, we
excluded raters who submitted more than $20\%$ of their ratings without playing
the audio. The SMOS study had $20$ raters, and the MOS study had $22$ raters
after exclusion.

\begin{table*}[ht]
	\begin{minipage}{0.60\textwidth}
		\centering \small
		\setlength{\tabcolsep}{3.5pt}
		\caption{Re-voicing EN-80}
		\label{tab:revoicing}
		\begin{tabular}{lrcccccc}
			\toprule
			Reference          & $n$   & Sim.           & Sim.              & Sim.    & DNSMOS     & NISQA      & WER / CER       \\
			                   &       & ref $\uparrow$ & src $ \downarrow$ & ref-src & $\uparrow$ & $\uparrow$ & $\downarrow$    \\
			\midrule
			Own utterance      & $160$ & --             & $.903$            & --      & $3.415$    & $4.618$    & $.095$ / $.052$ \\
			Same spk., other   & $160$ & $.869$         & $.806$            & $.817$  & $3.413$    & $4.610$    & $.099$ / $.055$ \\
			\midrule
			Different speaker  & $320$ & $.763$         & $.283$            & $.191$  & $3.374$    & $4.587$    & $.140$ / $.079$ \\
			\quad same gender  & $160$ & $.753$         & $.373$            & $.248$  & $3.397$    & $4.619$    & $.122$ / $.069$ \\
			\quad cross gender & $160$ & $.772$         & $.194$            & $.134$  & $3.350$    & $4.555$    & $.158$ / $.089$ \\
			\bottomrule
		\end{tabular}
	\end{minipage}
	\hfill
	\begin{minipage}{0.34\textwidth}
		\centering \small
		\setlength{\tabcolsep}{3.5pt}
		\caption{Speaker similarity, $20$ listeners}
		\label{tab:smos}
		\begin{tabular}{lcc}
			\toprule
			Condition           & SMOS   & 95\% CI        \\
			\midrule
			Same speaker        & $3.89$ & $[3.63, 4.15]$ \\
			\midrule
			Ref., matched lang. & $3.58$ & $[3.33, 3.83]$ \\
			Ref., cross lang.   & $3.25$ & $[2.99, 3.51]$ \\
			Src., matched lang. & $2.05$ & $[1.77, 2.33]$ \\
			Src., cross lang.   & $1.38$ & $[1.21, 1.55]$ \\
			\midrule
			Different speakers  & $1.50$ & $[1.30, 1.70]$ \\
			\bottomrule
		\end{tabular}
	\end{minipage}
\end{table*}

\section{Results}
\label{sec:results}

\textbf{The embedding controls speaker identity.} Identity is carried by the
conditioning embedding rather than the codes themselves.
Table~\ref{tab:revoicing} shows the results of decoding the same source codes
with an embedding from a reference speaker. Similarity to the reference of a
different speaker reaches $.763$, compared to $.869$ when the reference is a
different utterance from the same speaker. Similarity to the source drops to
$.283$ while the similarity between genuinely different speakers is at $.191$.
This shows that the source speaker signal is largely removed from the output.

Cosine similarity does not reflect the perceptual scale. So, we use perceptual
human evaluation (SMOS) to ground the cosine similarity. Table~\ref{tab:smos}
shows that re-voiced utterances are close to genuine same-speaker pairs ($3.58$
vs. $3.89$). When the reference is in another language, it is no more similar
to the source than genuinely different speakers ($1.38$ vs. $1.50$). A two-way
repeated-measures ANOVA over per-listener cell means shows that the comparison
effect is large ($F(1,19) = 78.7$, $p < .001$, $\eta_p^2 = .81$). This
indicates that the outputs are rated much closer to the reference speaker than
the source, confirming the effectiveness of our method.

\textbf{Reduced recoverable speaker information from tokens.} Standalone
speaker probe training shows that our proposed method reduces recoverable
speaker information by $30\%$: top-1 accuracy falls from $7.36\%$ to $5.14\%$
over $41{,}481$ classes against a chance rate of $0.002\%$. The unconditional
baseline is uniquely correct on 265 speakers, the conditional model on 154. The
differences between models are statistically significant according to paired
McNemar test~\cite{mcnemarNoteSamplingError1947} with $p<10^{-7}$. The relative
reduction rate holds equally on held-out utterances subset (n=355); from
$15.8\%$ to $11.0\%$ for unconditional baseline model and 21Codec respectively.

\begin{table}[t]
	\centering
	\small
	\setlength{\tabcolsep}{3.5pt}
	\caption{Decoder chunk size vs.\ quality for 21Codec.
		$\dagger$ marks chunk sizes seen during training}
	\label{tab:chunk}
	\begin{tabular}{lcccc}
		\toprule
		Chunk                                                  & Spk. sim.  & WER/CER      & DNSMOS     & NISQA      \\
		                                                       & $\uparrow$ & $\downarrow$ & $\uparrow$ & $\uparrow$ \\
		\midrule
		\phantom{00}1$^\dagger$ (\phantom{0}20\,ms)            & .854       & .107 / .062  & 3.28       & 4.58       \\
		\phantom{00}4$^\dagger$ (\phantom{0}80\,ms)            & .872       & .099 / .055  & 3.32       & 4.64       \\
		\phantom{00}8$^\dagger$ (160\,ms)                      & .875       & .095 / .052  & 3.34       & 4.68       \\
		\phantom{0}16$^\dagger$ (320\,ms)                      & .878       & .093 / .052  & 3.34       & 4.69       \\
		\phantom{0}25\phantom{0} (500\,ms)                     & .881       & .087 / .051  & 3.35       & 4.68       \\
		\phantom{0}50\phantom{0} (\phantom{00}1\,s\phantom{0}) & .882       & .097 / .056  & 3.36       & 4.69       \\
		\midrule
		non-causal                                             & .881       & .104 / .069  & 3.36       & 4.68       \\
		\bottomrule
	\end{tabular}
\end{table}

\begin{figure}[t]
	\begin{minipage}[b]{0.8\linewidth}
		\centering
		\centerline{\includegraphics[width=\linewidth]{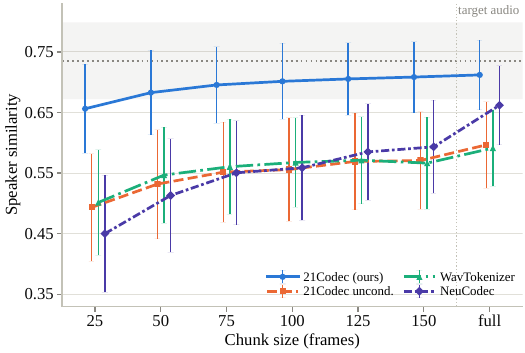}}
	\end{minipage}
	\caption{Speaker drift under \emph{un-cached} chunked decoding.
		Markers are means over 2\,s windows and files, bars the within-utterance standard deviation.
		The gray band is natural audio through the same windowing ($0.735 \pm 0.063$), the zero-drift floor.}
	\label{fig:drift}
\end{figure}

\textbf{Decoding in chunks does not change the voice.} We evaluate chunk sizes
from $20$ ms to $1$ s, including chunk sizes that were never seen during
training. Reconstruction quality is not affected, and the speaker similarity
remains within $.002$ of the non-causal decoder at $500$ ms
(Table~\ref{tab:chunk}). Our model is chunk-causal, so these measurements
reflect the real-time streaming setup where the decoder can use cache from
previous chunks. We also measure the real-time factor (RTF) of the proposed
model on CPU and GPU. Streaming decoding runs at RTF $.129$ on CPU and $.035$
on GPU at a 160 ms chunk (Intel i9-13900K, RTX 4090, batch 1, ONNX Runtime).
Per-call latency is nearly independent of chunk size ($16.8–20.6$ ms on CPU,
$5.3-5.5$ ms on GPU across $20-160$ ms chunks). Additionally, we run all the
models on uncached chunked decoding, where the same encoder output is decoded
in chunks without any cache from previous chunks (speaker drift evaluation).
Figure \ref{fig:drift} shows that our proposed model is able to maintain the
speaker identity across chunks, while other models drift to different speakers.

\begin{table}[t]
	\centering \small
	\setlength{\tabcolsep}{3.5pt}
	\caption{Reconstruction on EN-80, $n=160$.
		All systems run at $0.80$\,kbps except WavTokenizer ($0.90$).}
	\label{tab:reconstruction}
	\begin{tabular}{lcccc}
		\toprule
		System       & Sim.\ src  & DNSMOS     & NISQA      & WER / CER       \\
		             & $\uparrow$ & $\uparrow$ & $\uparrow$ & $\downarrow$    \\
		\midrule
		Ground Truth & --         & $3.390$    & $4.754$    & --              \\
		\midrule
		NeuCodec     & $.889$     & $3.309$    & $4.735$    & $.027$ / $.013$ \\
		WavTokenizer & $.801$     & $3.408$    & $4.698$    & $.037$ / $.019$ \\
		\midrule
		21Codec      & $.903$     & $3.415$    & $4.618$    & $.095$ / $.052$ \\
		21Codec-U    & $.818$     & $3.378$    & $4.569$    & $.047$ / $.025$ \\
		\bottomrule
	\end{tabular}
\end{table}

\textbf{The cost of an identity-free token stream.} Our proposed method
successfully improves the controllability of the speaker identity, but it comes
at a cost of intelligibility. Table~\ref{tab:reconstruction} reports
reconstruction on unseen speakers. Against the unconditional baseline, the
proposed model raises WER from $.047$ to $.095$ while improving speaker
similarity from $.818$ to $.903$. Naturalness, however, is not part of the
trade-off. DNSMOS and NISQA are higher for the conditional model ($3.415$ vs.\
$3.378$; $4.618$ vs.\ $4.569$), and $22$ listeners rate naturalness for the two
settings indistinguishably ($3.30$ $[3.10,3.50]$ vs. $3.19$ $[2.98, 3.39]$).
The listening test also separates the systems where the non-intrusive metrics
do not: DNSMOS places 21Codec first of five and above unprocessed audio, while
listeners place it fourth, behind WavTokenizer ($3.98$) and unprocessed audio
($3.92$). We therefore read DNSMOS and NISQA as evidence that conditioning does
not degrade quality, rather than as a ranking across codecs.

\section{Conclusions} \label{sec:conclusion}
We propose a streaming codec that suppresses the speaker identity in its token
stream and reintroduces it at the decoder side with an external identity
signal. Re-voiced output reaches a similarity of $.763$ to the unseen
reference, while the similarity to the source drops to $.283$. Our findings
show the potential of decoder-side speaker identity control and the trade-off
between the intelligibility and speaker removal from the token stream.

\textbf{Limitations and future work.} Training was stopped after 5 days for
budget reasons, and had not fully converged. So, reported numbers are lower
bound. The method also depends on the speaker embedding model; a different
embedding model or a learned speaker table may shift the results.

\bibliographystyle{IEEEbib} \bibliography{bib/refs}

\end{document}